# A Preliminary Mathematical Model for the Dynamic Transmission of Dengue, Chikungunya and Zika

Raúl Isea[1, *], Karl E. Lonngren[2]

[1]Institute of Advanced Studies – IDEA, Hoyo de la Puerta, Baruta, Venezuela

[2]Department of Electrical and Computer Engineering, University of Iowa, Iowa City, Iowa, USA

### Email address

risea@idea.gob.ve (R. Isea)

[*]Corresponding author

### Abstract

*Aedes aegypti* is a known vector of Dengue, Chikungunya and Zika and the goal of this study is to propose the first mathematical model to describe the dynamic transmission of these three diseases. We present two preliminary models that consist of the SEIR model for the human populations and an SEI model for the vector to describe (a) the single transmission dynamics of dengue, Chikungunya or Zika, and (b) any possible coinfection between two diseases in the same population. In order to do that, we obtain an analytical solution of the system of 17 and 30 coupled differential equations for each model respectively, and later obtain the eigenvalues by analyzing the Jacobian matrix in order to begin the development of a surveillance system to prevent the spread of these three diseases.

## 1. Introduction

Recently, cases of Dengue, Chikungunya and Zika have been confirmed in Africa, Southeast Asia, the Pacific Islands, the Caribbean and Latin America and unfortunately, a vaccine or an effective treatment is not currently available. All diseases were transmitted through the bite of the female mosquito *Aedes aegypti* [1, 2].

Zika is an emerging mosquito-borne virus genus *Flavivirus* and it was first isolated in Uganda [3] and it has been confirmed to exist recently in the United Kingdom, the United States, Southeast Asia, the Pacific Islands, and South America. In parallel, it has been reported that a possible coinfection could occur between these mosquito transmitted diseases. For example, Dengue and Zika occurred in two patients in New Caledonia [4], Dengue and Chikungunya occurred in Central Africa where 1567 patients tested positive for Chikungunya, 376 for Dengue serotype 2 and 37 were coinfected with both viruses [5]. Just recently, it was reported that a Colombian patient was coinfected with all three diseases in 2016 [6].

It is necessary to develop surveillance systems to prevent the spread of these three epidemics. Up to the present time, a mathematical model that considers the transmission dynamics of a triple epidemic outbreak has not been proposed.

We propose two mathematical models to explain the transmission dynamic of Dengue, Chikungunya and Zika by employing the SEIR/SEI models.

## 2. Mathematical Model

The models are based on the recent model proposed by Isea [7]. In this model, the host population ($N$) is subdivided into multiple subpopulations based on the following criteria:

- The model assumes that a human cannot infect another human.

- The model excludes the transmission of Zika from person to person by sexual contact because the percentage of infected people is very small in comparison with Chikungunya and Dengue patients.

- The model only considers one subtype of Dengue.

- The total mosquito (vector) population is denoted by $M$

and it is divided into 7 classes. The first is $S'_v$ which represents the mosquito population carrying the virus. The next four are the mosquitoes that are exposed to the Dengue, Chikungunya and Zika viruses. We write

$$M = S'_v + E'_{vD} + E'_{vC} + E'_{vZ} + I'_{vD} + I'_{vC} + I'_{vZ} \quad (1)$$

where the prime indicates the derivative with respect to time.

The next section will present the two models that are proposed in this work.

MODEL 1

The initial model considers that the total human population at time *t* is divided into 10 subpopulations. This model does not consider coinfection between Dengue and Chikungunya, Chikungunya and Zika, or any other possible combinations between them and only later, coinfection between them is considered.

The susceptible population ($S_h$) that will be exposed to an infection by one type of virus is denoted by ($E_D$, $E_C$, $E_Z$) where the subscript 'D', 'C' and 'Z' refer to Dengue, Chikungunya and Zika, respectively. Subsequently, the populations will become infected and it is denoted by ($I_D$, $I_C$, $I_Z$). Finally the human population that recovers is indicated with ($R_D$, $R_C$, $R_Z$). The total population (N) is given by:

$$N = S'_h + E'_D + E'_C + E'_Z + I'_D + I'_C + I'_Z + R'_D + R'_C + R'_Z \quad (2)$$

where the prime indicates derivatives with respect to time *t*.

The model 1 consists of a set of 17 ordinary differential equations:

$$\frac{dS_h}{dt} = \mu(N - S_h) - \frac{S_h}{N}(\beta_D I_{vD} + \beta_C I_{vC} + \beta_Z I_{vZ}) \quad (3)$$

$$\frac{dE_i}{dt} = \frac{S_h}{N}[\beta_i I_{vi} - (\sigma_i + \mu) E_i] \quad (4)$$

$$\frac{dI_i}{dt} = \sigma_i E_i - (\gamma_i + \mu) I_i \quad (5)$$

$$\frac{dR_i}{dt} = \sum_{i=1}^{3}[\gamma_i I_i - \mu R_i] \quad (6)$$

$$\frac{dS_v}{dt} = M - S_v \frac{\beta_v}{N}(I_D + I_C + I_Z) - \mu_v S_v \quad (7)$$

$$\frac{dE_{vi}}{dt} = S_v \frac{\beta_v}{N} I_i - (\sigma_v + \mu_v) E_{vi} \quad (8)$$

$$\frac{dI_{vi}}{dt} = \sigma_v E_{vi} - (\gamma_v + \mu_v) I_{vi} \quad (9)$$

where *i* = 1, 2, 3 which represent 'D', 'C' and 'Z', respectively. We assumed that the birth and the death rates are equal and are denoted by $\mu$; ($\gamma_D$, $\gamma_C$, $\gamma_Z$) are the mean infectious periods; ($\sigma_D$, $\sigma_C$, $\sigma_Z$) are the mean latent periods; and ($\beta_D$, $\beta_C$, $\beta_Z$) are the transmission parameters of Dengue, Chikungunya and Zika, respectively.

MODEL 2

This model only considers that one person will be infected by one disease and be later coinfected by the other diseases. Suppose that one is initially exposed to Dengue ($E_D$) and eventually becomes infected with this disease ($I_D$) and can or cannot recover ($R_D$). Later, this same individual is exposed to Chickungunya or Zika, and we represent it with two subindices: *DC* or *DZ* that represents that it was exposed to Dengue and Chikungunya, or Dengue and Zika, respectively. Figure 1 shows the SEIR model of population employed in this work.

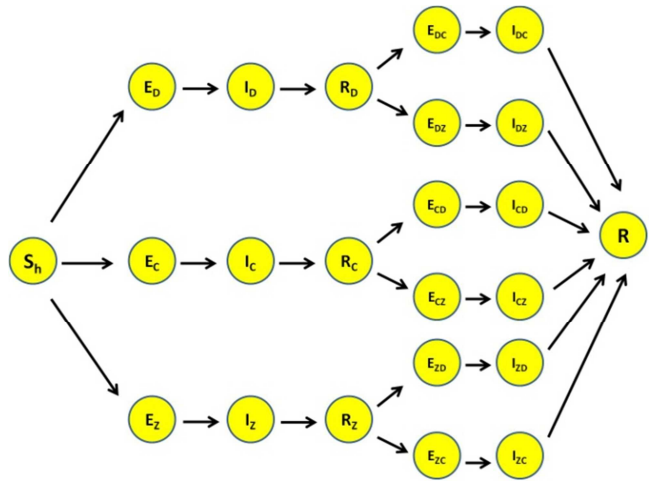

*Figure 1.* The compartmental diagram of the host population employed in the Model 2.

Therefore the differential equations of Model 2 are 30 equations (see equations 10 until 19):

$$\frac{dS_h}{dt} = \mu(N - S_h) - \frac{S_h}{N}(\beta_{vD} I_{vD} + \beta_{vC} I_{vC} + \beta_{vZ} I_{vZ}) \quad (10)$$

$$\frac{dE_i}{dt} = \frac{S_h}{N}[\beta_{vi} I_{vi} - (\sigma_i + \mu) E_i] \quad (11)$$

$$\frac{dI_i}{dt} = \sigma_i E_i - (\gamma_i + \mu) I_i \quad (12)$$

$$\frac{dR_i}{dt} = \gamma_i I_i - \mu R_i - \frac{R_i}{N} \sum_{\substack{j=1 \\ j \neq i}}^{3} \beta_{vi} I_{vi} \quad (13)$$

$$\frac{dE_{ij}}{dt} = R_i \frac{\beta_{vj}}{N} I_{vj} - (\sigma_j + \mu) E_{ij} \quad (14)$$

$$\frac{dI_{ij}}{dt} = \sigma_j E_{ij} - (\gamma_j + \mu) I_{ij} \quad (15)$$

$$\frac{dR}{dt} = \sum_{\substack{i=1 \\ i \neq j}}^{3} \gamma_i I_{ji} - \mu R \quad (16)$$

$$\frac{dS_v}{dt} = M - S_v \frac{\beta_v}{N}(I_D + I_C + I_Z) - \mu_v S_v \quad (17)$$

$$\frac{dE_{vi}}{dt} = S_v \frac{\beta_v}{N} I_i - (\sigma_v + \mu_v) E_{vi} \quad (18)$$

$$\frac{dI_{vi}}{dt} = [\sigma_v E_{vi} - (\gamma_v + \mu_v) I_{vi}] \quad (19)$$

where *i* = 1, 2, 3 which represent 'D', 'C' and 'Z', respectively. The term $E_{ij}$ corresponds to the following terms $E_{DC}$, $E_{DZ}$, $E_{CD}$, $E_{CZ}$, $E_{ZC}$, $E_{ZD}$; while $I_{ij}$ corresponds to $I_{DC}$, $I_{DZ}$, $I_{CD}$, $I_{CZ}$, $I_{ZC}$, $I_{ZD}$.

## 3. Results

The epidemiologically relevant bioregion that is symbolized with $\Omega$ [7, 8] is given by

*Model 1:*

$$\Omega_1 = (S'_h, E'_D, E'_C, E'_Z, I'_D, I'_C, I'_Z, R'_D, R'_C, R'_Z, S'_v, E'_{vD}, E'_{vC}, E'_{vZ}, I'_{vD}, I'_{vC}, I'_{vZ}) \quad (20)$$

*Model 2:*

$$\Omega_2 = (S'_h, E'_D, E'_C, E'_Z, I'_D, I'_C, I'_Z, R'_D, R'_C, R'_Z, E'_{DC}, E'_{DZ}, E'_{CD}, E'_{CZ}, E'_{ZD}, E'_{ZC}, I'_{DC}, I'_{DZ}, I'_{CD}, I'_{CZ}, I'_{ZD}, I'_{ZC}, R', S'_v, E'_{vD}, E'_{vC}, E'_{vZ}, I'_{vD}, I'_{vC}, I'_{vZ}) \quad (21)$$

The solution of this system of equations uses the same methodology as explained in Isea [7], Janreug and Chrrivirigasit [8], and employed previously [9-11]. We show the results for each model.

*Model 1:*
We obtain a critical point of the system of equations (3-9):

$$S_h = \frac{\mu N}{\mu + C_1} \quad (22)$$

$$E_D = \frac{\mu \beta_D I_{vD}}{(\mu + \sigma_D)(C_1 + \mu)} \quad (23)$$

$$E_C = \frac{\mu \beta_C I_{vC}}{(\mu + \sigma_C)(C_1 + \mu)} \quad (24)$$

$$E_Z = \frac{\mu \beta_Z I_{vZ}}{(\mu + \sigma_Z)(C_1 + \mu)} \quad (25)$$

$$I_D = \frac{\beta_D \mu \sigma_D I_{vD}}{C_1(1+\mu)[\sigma_D(\gamma_D+\mu)+\mu\gamma_D+\mu^2]} \quad (26)$$

$$I_C = \frac{\beta_C \mu \sigma_C I_{vC}}{C_1(1+\mu)[\sigma_C(\gamma_C+\mu)+\mu\gamma_C+\mu^2]} \quad (27)$$

$$I_Z = \frac{\beta_Z \mu \sigma_Z I_{vZ}}{C_1(1+\mu)[\sigma_Z(\gamma_Z+\mu)+\mu\gamma_Z+\mu^2]} \quad (28)$$

$$S_v = \frac{M N}{C_2} \quad (29)$$

$$E_{vD} = \frac{M \beta_{vD} I_D}{N C_2 C_3} \quad (30)$$

$$E_{vC} = \frac{M \beta_{vC} I_C}{N C_2 C_3} \quad (31)$$

$$E_{vZ} = \frac{M \beta_{vZ} I_Z}{N C_2 C_3} \quad (32)$$

$$I_{vD} = \frac{M \sigma_v \beta_v I_D}{C_2 C_3^2} \quad (33)$$

$$I_{vC} = \frac{M \sigma_v \beta_v I_C}{C_2 C_3^2} \quad (34)$$

$$I_{vZ} = \frac{M \sigma_v \beta_v I_Z}{C_2 C_3^2} \quad (35)$$

where the constants are:

$$C_1 \equiv \mu N + \beta_D I_{vD} + \beta_C I_{vC} + \beta_Z I_{vZ}$$

$$C_2 \equiv \mu_v N + \beta_v (I_D + I_C + I_Z)$$

$$C_3 \equiv \sigma_v - \mu_v$$

The next step is to obtain the Jacobian matrix denoted by J (equation 36), *ie.* the partial derivative of the differential equations evaluated at the critical point. The elements of the Jacobian are:

$$J = \begin{pmatrix} J_{1,1} & \cdots & J_{1,17} \\ \vdots & \vdots & \vdots \\ J_{17,1} & \cdots & J_{17,17} \end{pmatrix} \quad (36)$$

The non-zero elements are:

$$J_{1,1} = -\frac{1}{N}(\beta_{vD} I_{vD} + \beta_{vC} I_{vC} + \beta_{vZ} I_{vZ}) - \mu;$$

$$J_{2,1} = \frac{\beta_{vD} I_{vD}}{N}; J_{2,2} = -\sigma_D - \mu; J_{3,1} = \frac{\beta_{vC} I_{vC}}{N};$$

$$J_{3,3} = -\sigma_C - \mu; J_{4,1} = \frac{\beta_{vZ} I_{vZ}}{N}; J_{4,4} = -\sigma_Z - \mu;$$

$$J_{5,2} = -\sigma_D; J_{5,5} = -\sigma_D - \mu; J_{6,3} = -\sigma_C;$$

$$J_{6,6} = -\sigma_C - \mu;$$

$$J_{7,4} = -\sigma_Z; J_{7,7} = -\sigma_Z - \mu; J_{8,5} = \gamma_D; J_{9,6} = \gamma_C;$$

$$J_{8,8} = J_{9,9} = J_{10,10} = -\mu; J_{10,7} = \gamma_Z; J_{12,11} = \frac{\beta I_D}{N};$$

$$J_{11,11} = -\frac{\beta}{N}(I_D + I_C + I_Z) - \mu_v; J_{13,11} = \frac{\beta I_C}{N};$$

$$J_{14,11} = \frac{\beta I_Z}{N}; J_{12,12} = J_{13,13} = J_{14,14} = -\sigma_v - \mu_v;$$

$$J_{15,12} = J_{16,13} = J_{17,14} = \gamma_v;$$

$$J_{15,15} = J_{16,16} = J_{17,17} = -\gamma_v - \mu_v;$$

The determinant of Jacobian matrix is equal to:

$$-\mu^3 (-\mu_v - \gamma_v)^3 (-\sigma_v - \mu_v)^3 (-\gamma_C - \mu)(-\gamma_D - \mu)$$
$$(-\gamma_Z - \mu)(-\sigma_C - \mu)(-\sigma_D - \mu)(-\sigma_Z - \mu)$$
$$\left(-\frac{\beta_{vD} I_{vD}}{N} - \frac{\beta_{vC} I_{vC}}{N} - \frac{\beta_{vZ} I_{vZ}}{N} - \mu\right)$$
$$\left(-\frac{\beta I_D}{N} - \frac{\beta I_C}{N} - \frac{\beta I_Z}{N} - \mu_v\right)$$

Finally, we obtained the eigenvalues this by examining the Jacobian evaluated at the critical point. We obtain:

$$-\sigma_d - \mu; \; -\sigma_c - \mu; \; -\sigma_z - \mu;$$

$$\gamma_d - \mu; \; \gamma_c - \mu; \; \gamma_z - \mu$$

the solution is stable when this eigenvalues are negative, and for this reason, we find that $\gamma_d < \mu$, $\gamma_c < \mu$ and $\gamma_z < \mu$,

*Model 2:*
We seek a one nontrivial point of equilibrium of the system of equations and find:

$$S_h = \frac{\mu N}{\mu + F} \quad (37)$$

$$E_D = \frac{\mu \beta_{vD} I_{vD}}{(\mu + \sigma_D)(F + \mu) + \mu^2} \quad (38)$$

$$E_C = \frac{\mu \beta_{vC} I_{vC}}{(\mu + \sigma_C)(F + \mu) + \mu^2} \quad (39)$$

$$E_Z = \frac{\mu \beta_{vZ} I_{vZ}}{(\mu + \sigma_Z)(F + \mu) + \mu^2} \quad (40)$$

$$I_D = \frac{\beta_{vD}\,\mu\,\sigma_D\,I_{vD}}{(\sigma_D+\mu)\,(F+\mu)\,(\gamma_D+\mu)} \quad (41)$$

$$I_C = \frac{\beta_{vC}\,\mu\,\sigma_C\,I_{vC}}{(\sigma_C+\mu)\,(F+\mu)\,(\gamma_C+\mu)} \quad (42)$$

$$I_Z = \frac{\beta_{vZ}\,\mu\,\sigma_Z\,I_{vZ}}{(\sigma_Z+\mu)\,(F+\mu)\,(\gamma_Z+\mu)} \quad (43)$$

$$R_D = \frac{\mu\,N\,\beta_{vD}\,I_{vD}}{(\beta_{vC}I_{vC}+\beta_{vZ}I_{vZ})\,(F+2\mu)+\mu(\beta_{vD}I_{vD}+\mu N)} \quad (44)$$

$$R_C = \frac{\mu\,N\,\beta_{vC}\,I_{vC}\,\sigma_C\,\gamma_C}{(\beta_{vD}I_{vD}+\beta_{vZ}I_{vZ})\,(F+2\mu)+\mu(\beta_{vC}I_{vC}+\mu N)} \quad (45)$$

$$R_Z = \frac{\mu\,N\,\beta_{vZ}\,I_{vZ}\,\sigma_Z\,\gamma_Z}{(\beta_{vC}I_{vC}+\beta_{vD}I_{vD})\,(F+2\mu)+\mu(\beta_{vZ}I_{vZ}+\mu N)} \quad (46)$$

$$E_{DC} = \frac{\mu\,\beta_{vD}\,I_{vD}\,\beta_{vC}\,I_{vC}\,\gamma_D\,\sigma_D}{(\beta_{vC}I_{vC}+\beta_{vZ}I_{vZ})\,(F+2\mu)+\mu(\beta_{vD}I_{vD}+\mu N)} \quad (47)$$

$$E_{DZ} = \frac{\mu\,\beta_{vD}\,I_{vD}\,\beta_{vZ}\,I_{vZ}\,\sigma_D\,\gamma_D}{(\beta_{vC}I_{vC}+\beta_{vZ}I_{vZ})\,(F+2\mu)+\mu(\beta_{vD}I_{vD}+\mu N)} \quad (48)$$

$$E_{CD} = \frac{\mu\,\beta_{vC}\,I_{vC}\,\beta_{vD}\,I_{vD}\,\sigma_C\,\gamma_C}{(\beta_{vD}I_{vD}+\beta_{vZ}I_{vZ})\,(F+2\mu)+\mu(\beta_{vC}I_{vC}+\mu N)} \quad (49)$$

$$E_{CZ} = \frac{\mu\,\beta_{vC}\,I_{vC}\,\beta_{vZ}\,I_{vZ}\,\sigma_C\gamma_C}{(\beta_{vD}I_{vD}+\beta_{vZ}I_{vZ})\,(F+2\mu)+\mu(\beta_{vC}I_{vC}+\mu N)} \quad (50)$$

$$E_{ZC} = \frac{\mu\,\beta_{vZ}\,I_{vZ}\,\beta_{vC}\,I_{vC}\,\sigma_Z\,\gamma_Z}{(\beta_{vD}I_{vD}+\beta_{vC}I_{vC})\,(F+2\mu)+\mu(\beta_{vZ}I_{vZ}+\mu N)} \quad (51)$$

$$E_{ZD} = \frac{\mu\,\beta_{vZ}\,I_{vZ}\,\beta_{vD}\,I_{vD}\,\sigma_Z\,\gamma_Z}{(\beta_{vD}I_{vD}+\beta_{vC}I_{vC})\,(F+2\mu)+\mu(\beta_{vZ}I_{vZ}+\mu N)} \quad (52)$$

$$I_{DC} = \frac{\mu\,\beta_{vD}\,I_{vD}\,\beta_{vC}\,I_{vC}\,\gamma_D\,\sigma_D\,\sigma_C}{(\beta_{vC}I_{vC}+\beta_{vZ}I_{vZ})\,(F+2\mu)+\mu(\beta_{vD}I_{vD}+\mu N)} \quad (53)$$

$$I_{DZ} = \frac{\mu\,\beta_{vD}\,I_{vD}\,\beta_{vZ}\,I_{vZ}\,\sigma_D\,\gamma_D\,\sigma_Z}{(\beta_{vC}I_{vC}+\beta_{vZ}I_{vZ})\,(F+2\mu)+\mu(\beta_{vD}I_{vD}+\mu N)} \quad (54)$$

$$I_{CD} = \frac{\mu\,\beta_{vC}\,I_{vC}\,\beta_{vD}\,I_{vD}\,\sigma_C\,\gamma_C\,\sigma_D}{(\beta_{vD}I_{vD}+\beta_{vZ}I_{vZ})\,(F+2\mu)+\mu(\beta_{vC}I_{vC}+\mu N)} \quad (55)$$

$$I_{CZ} = \frac{\mu\,\beta_{vC}\,I_{vC}\,\beta_{vZ}\,I_{vZ}\,\sigma_C\,\gamma_C\,\sigma_Z}{(\beta_{vD}I_{vD}+\beta_{vZ}I_{vZ})\,(F+2\mu)+\mu(\beta_{vC}I_{vC}+\mu N)} \quad (56)$$

$$I_{ZC} = \frac{\mu\,\beta_{vZ}\,I_{vZ}\,\beta_{vC}\,I_{vC}\,\sigma_Z\,\gamma_Z\,\sigma_C}{(\beta_{vD}I_{vD}+\beta_{vC}I_{vC})\,(F+2\mu)+\mu(\beta_{vZ}I_{vZ}+\mu N)} \quad (57)$$

$$I_{ZD} = \frac{\mu\,\beta_{vZ}\,I_{vZ}\,\beta_{vD}\,I_{vD}\,\sigma_Z\,\gamma_Z\,\sigma_D}{(\beta_{vD}I_{vD}+\beta_{vC}I_{vC})\,(F+2\mu)+\mu(\beta_{vZ}I_{vZ}+\mu N)} \quad (58)$$

$$S_v = \frac{M}{A+\mu} \quad (59)$$

$$E_{vi} = \sum_{i=1}^{3} \frac{M\,\beta_v\,I_i}{N\,(\gamma_v+\mu_v)\,(A+\mu_v)} \quad (60)$$

$$I_{vi} = \sum_{i=1}^{3} \frac{M\,\sigma_v\,\beta_v\,I_i}{N\,(A+\mu_v)\,(\gamma_v+\mu_v)\,(\sigma_v+\mu_v)} \quad (61)$$

where $i$ correspond to 1 until 3 for Dengue, Chikungunya and Zika, respectively; and we defined the following variables:

$$BD \equiv \frac{\beta_{vD}\,I_{vD}}{N};\ BC \equiv \frac{\beta_{vC}\,I_{vC}}{N};\ BZ \equiv \frac{\beta_{vZ}\,I_{vZ}}{N};$$

$$F \equiv BC + BD + BZ$$

$$A \equiv \frac{\beta_v}{N}\,(I_C + I_D + I_Z)$$

Similarly with the Model 1, they are 63 nonzero elements of Jacobian matrix, but only show the first 32:

$$J_{1,1} = -\frac{1}{N}\,(\beta_{vD}I_{vD} + \beta_{vC}I_{vC} + \beta_{vZ}I_{vZ}) - \mu;$$

$$J_{2,1} = \frac{\beta_{vD}I_{vD}}{N};\ J_{2,2} = -\sigma_D - \mu;\ J_{3,1} = \frac{\beta_{vC}I_{vC}}{N};$$

$$J_{3,3} = J_{11,11} = -\sigma_C - \mu;\ J_{4,1} = \frac{\beta_{vZ}I_{vZ}}{N};$$

$$J_{4,4} = J_{12,12} = -\sigma_Z - \mu;$$

$$J_{5,2} = -\sigma_D;\ J_{5,5} = J_{13,13} = -\sigma_D - \mu;\ J_{6,3} = -\sigma_C;$$

$$J_{6,6} = -\sigma_C - \mu;$$

$$J_{7,4} = -\sigma_Z;\ J_{7,7} = -\sigma_Z - \mu;\ J_{8,5} = \gamma_D;\ J_{9,6} = \gamma_C;$$

$$J_{8,8} = -\frac{1}{N}\,(\beta_{vC}I_{vC} + \beta_{vZ}I_{vZ}) - \mu;\ J_{10,7} = \gamma_Z;$$

$$J_{9,9} = -\frac{1}{N}\,(\beta_{vD}I_{vD} + \beta_{vZ}I_{vZ}) - \mu;$$

$$J_{10,10} = -\frac{1}{N}\,(\beta_{vD}I_{vD} + \beta_{vC}I_{vC}) - \mu;$$

$$J_{10,7} = \gamma_Z;\ J_{11,8} = \frac{\beta\,I_{vC}}{N};\ J_{12,8} = \frac{\beta\,I_{vZ}}{N};\ J_{12,8} = \frac{\beta\,I_{vD}}{N};$$

$$J_{11,11} = -\frac{\beta}{N}(I_D + I_C + I_Z) - \mu_v;\ J_{13,11} = \frac{\beta I_C}{N};$$

$$J_{14,11} = \frac{\beta I_Z}{N};\ J_{12,12} = J_{13,13} = J_{14,14} = -\sigma_v - \mu_v;$$

For this model, it is complicated to obtain the eigenvalues and it is necessary to dedicate more time for a complete understanding. In addition, the most critical value in the epidemic model is obtaining the Basic Reproduction Value ($R_0$), but this result is difficult to perform and will be analyzed in the future.

However, it is possible to resolve numerically the equations in the Model 1 using the Python program (see Figure 2), where we have employed an arbitrary choice for the parameters. For this example, we have assumed 3 cases of Dengue, 4 of Chikungunya and 1 for Zika.

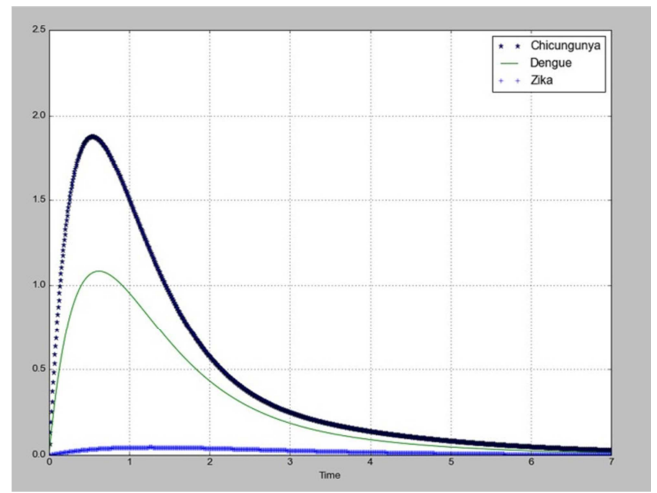

***Figure 2.*** *Numerical solution of the Model 1 presented with 3 cases of Dengue, 4 Chikungunya (asterisk) and 1 for Zika 1 versus time. The parameters were selected randomly.*

## 4. Conclusion

We have developed a preliminary model for the transmission dynamics of a triple epidemic outbreak employing the SEIR-SEI model. The initial models reveal that the critical points are not trivial to obtain and it is necessary to expand in this subject in the future, but we believe that these equations may initiate such a study. This model is the first development to help to prevent outbreaks from these epidemics, and the next step in the future is to determine the basic reproductive number $R_0$ which is the number of secondary cases which one case would produce in a complete susceptible population, and adjust the parameters of the mathematical model with the cases reported by country.

## Acknowledgements

The comments obtained by members of the Iberoamerindian catedra of the bio-genocultural soveragnity.